# Comment on "Gain-assisted Superluminal Light Propagation"

Harry I. Ringermacher
*General Electric Research Center, Schenectady, NY 12301*

Lawrence R. Mead
*Dept. of Physics and Astronomy, University of Southern Mississippi
Hattiesburg, MS  39406-5046*

Abstract. -  Wang, Kuzmich and Dogariu, in their 20 July, 2000, Nature article, describe an experiment ostensibly measuring superluminal speeds of 310c via a few percent shift in time of an optical pulse undergoing anomalous dispersion in a pumped medium. This paper seems to have become part of the "superluminal lore" and is often referenced as a profound example of superluminality. We have closely examined the article and find serious flaws that need to be addressed.

Wang, Kuzmich and Dogariu, in their 20 July, 2000, Nature article[1], describe an experiment ostensibly measuring superluminal speeds of 310C via a 3.35% shift in time of an " almost no change in pulse shape" optical pulse.  Putting aside their detailed theoretical exposition of anomalous dispersion, which is undoubtedly correct, we are left with asking some rather simple questions:  What are they actually measuring?  What is the interpretation of what they measure? Unfortunately, their experiment leaves open many questions. "Almost" is not good enough when measuring 3% effects.  Although they emphasize that they used low intensity laser probe pulses to minimize saturation effects, they do not include a standard linearity evaluation covering their optical regime.  We have to take their word on this.  A 3% pulse shape distortion, measured any number of ways, could account for their superluminality. For example, we note that the resonant pulse has wiggles on top receding into the trailing edge. That undoubtedly affects the moments of the waveform and therefore the location of leading and trailing edges.  This signal distortion was not even addressed. What was presented is even more enigmatic.  They claim, in Fig. 4 showing the actual pulse waveform, that the 62 ns shift in the leading edge between resonant and reference pulses is equal to the shift in the trailing edge. Were that true, it would vindicate the no distortion claim. But their insets to the figure clearly show, even to the "untrained eye", that they are not equal and roughly vary by perhaps a factor of two on average. Indeed, it is not even stated if they took an average. Furthermore, the entry of the leading and trailing edges into the base of the waveform appear to be more consistent with that produced by a distortion rather than a simple shift in time as can be demonstrated by a theoretical calculation of a 3% shift for a "near Gaussian" pulse.

Even more disturbing is their defense of the Nature data on their internet site [2] using what is clearly different data.  In particular, the pulse, appearing in the internet data to be pure Gaussian with no wiggles, is not the same as the pulse presented in Nature. The wavefront edge insets on the website now appear to be ideal with equal leading and trailing time shifts of the Gaussian pulses. We do not know whether this is what they should have seen or what they did see and did not publish. A deeper understanding and exposition of effects of anomalous dispersion on waveform shape might have helped, but in principle is not essential – a better experiment and analysis is.  A shorter pulse that could be fully contained within the 6-cm cavity, for example, would have produced clearer results. Stating that causality is not violated by a pulse leaving the cavity before entering doesn't help either.

1. L.J. Wang, A. Kuzmich and A. Dogariu, Gain-assisted superluminal light propagation. Nature 406, 277-279 (2000)
2. www.neci.nj.nec.com/homepages/lwan/gas.htm